\documentclass[%
 reprint,
 bibnotes,
 amsmath,amssymb,
 aps,
 prl,
 ]{revtex4-1}

\usepackage{graphicx}
\usepackage{dcolumn}
\usepackage{bm}
\usepackage[pdfstartview=FitH,
CJKbookmarks=true,
bookmarksnumbered=true,
bookmarksopen=true,
colorlinks,
linkcolor=blue,
anchorcolor=blue,
citecolor=blue,
allcolors=blue,
pdfborder=001,
]{hyperref}
\usepackage[caption=false]{subfig}
\usepackage{floatrow}
\usepackage{enumitem}
\DeclareSubrefFormat{myparens}{#1~(#2)}
\DeclareCaptionListOfFormat{myparens}{#1(#2)}
\begin{document}

\title{\boldmath Determination of spin-parity
	quantum numbers of X(2370) as $0^{-+}$ from $J/\psi\rightarrow\gamma K^{0}_{S}K^{0}_{S}\eta^{\prime}$}

\author{
	M.~Ablikim$^{1}$, M.~N.~Achasov$^{4,b}$, P.~Adlarson$^{75}$, X.~C.~Ai$^{80}$, R.~Aliberti$^{35}$, A.~Amoroso$^{74A,74C}$, M.~R.~An$^{39}$, Q.~An$^{71,58}$, Y.~Bai$^{57}$, O.~Bakina$^{36}$
% ...
\\
\vspace{0.2cm}
(BESIII Collaboration)\\
\vspace{0.2cm} {\it
$^{1}$ Institute of High Energy Physics, Beijing 100049, People's Republic of China\\
% ...
$^{4}$ Budker Institute of Nuclear Physics SB RAS (BINP), Novosibirsk 630090, Russia\\
% ...
$^{35}$ Johannes Gutenberg University of Mainz, Johann-Joachim-Becher-Weg 45, D-55099 Mainz, Germany\\
$^{36}$ Joint Institute for Nuclear Research, 141980 Dubna, Moscow region, Russia\\
% ...
$^{39}$ Liaoning Normal University, Dalian 116029, People's Republic of China\\
% ...
$^{57}$ Southeast University, Nanjing 211100, People's Republic of China\\
$^{58}$ State Key Laboratory of Particle Detection and Electronics, Beijing 100049, Hefei 230026, People's Republic of China\\
% ...
$^{71}$ University of Science and Technology of China, Hefei 230026, People's Republic of China\\
% ...
$^{74}$ University of Turin and INFN, (A)University of Turin, I-10125, Turin, Italy; (B)University of Eastern Piedmont, I-15121, Alessandria, Italy; (C)INFN, I-10125, Turin, Italy\\
$^{75}$ Uppsala University, Box 516, SE-75120 Uppsala, Sweden\\
% ...
$^{80}$ Zhengzhou University, Zhengzhou 450001, People's Republic of China\\
\vspace{0.2cm}
% ...
$^{b}$ Also at the Novosibirsk State University, Novosibirsk, 630090, Russia\\
% ...
	}
}


{
\renewcommand{\baselinestretch}{1.20}

\begin{abstract}

Based on $(10087\pm44)\times10^{6}$ $J/\psi$ events collected with the BESIII detector, a partial wave analysis of the decay $J/\psi\rightarrow\gamma K^{0}_{S}K^{0}_{S}\eta^{\prime}$ is performed. 
The mass and width of the $X(2370)$ are measured to be $2395 \pm 11 ({\rm stat})^{+26}_{-94}({\rm syst})\ \mathrm{MeV}/c^{2}$ and $188^{+18}_{-17}({\rm stat})^{+124}_{-33}({\rm syst})~\mathrm{MeV}$, respectively. 
The corresponding product branching fraction is $\mathcal{B}[J/\psi\rightarrow\gamma X(2370)] \times \mathcal{B}[X(2370) \rightarrow f_{0}(980)\eta^{\prime}] \times \mathcal{B}[f_{0}(980) \rightarrow K^{0}_{S}K^{0}_{S}] = \left( 1.31 \pm 0.22 ({\rm stat})^{+2.85}_{-0.84}({\rm syst}) \right) \times 10^{-5}$. The statistical significance of the $X(2370)$ is greater than $11.7\sigma$ and the spin-parity is determined to be $0^{-+}$ for the first time. The measured mass and spin-parity of the $X(2370)$ are consistent with the predictions of the lightest pseudoscalar glueball.

\end{abstract}
\vspace{-2.0mm}
}
\maketitle

The non-Abelian property of Quantum Chromodynamics (QCD) permits the existence of new types of hadrons, such as glueballs, hybrids and multi-quark states, 
which are beyond conventional mesons and baryons in the constituent quark model \cite{AMSLER200461,KLEMPT20071,CREDE200974}.
In particular, the glueball is a unique particle formed via the interaction among gauge boson particles.
Lattice Quantum Chromodynamics (LQCD) predicts that the ground state of a pseudoscalar glueball has a mass around $2.3 - 2.6~\mathrm{GeV}$/${c}^{2}$ \cite{LQCD1,LQCD2,LQCD3,LQCD4,LQCD5}. 	
The radiative decay of the $J/\psi$ meson is a gluon-rich process and is therefore regarded as an ideal place for searching and studying glueballs \cite{PhysRevD.50.3268,PhysRevD.55.5749}. 

A $\pi^+\pi^-\eta^{\prime}$ resonance, the $X(2370)$, was observed in $J/\psi\rightarrow\gamma\pi^{+}\pi^{-}\eta^{\prime}$ with a statistical significance greater than $6.4\sigma$ in the BESIII experiment\cite{1835_confirmed}. It was further observed from the combined measurement of $J/\psi\rightarrow \gamma K^{+}K^{-}\eta^{\prime}$ and $J/\psi\rightarrow \gamma K^{0}_{S}K^{0}_{S}\eta^{\prime}$ with a statistical significance of $8.3\sigma$ by BESIII \cite{epjc.s10052_gongli}. 
This experimental observation stimulated a number of theoretical speculations \cite{2370_prediction_eta,2370_prediction_hexquark_2012,2370_prediction_baryonium_2012,2370_prediction_eta2_2020,2370_prediction_teraquark_2022} for its nature. 
Among them, one of the intriguing explanations is a pseudoscalar glueball \cite{LQCD5,2370_prediction_glueball_mix,PhysRevD.87.054036,Zhang:2021xvl}.
A high-statistics  $J/\psi$  data sample collected with BESIII provides an opportunity to further investigate the
properties of the $X(2370)$ and helps to understand the dynamics of QCD.

To understand the nature of the $X(2370)$, it is crucial to measure its quantum numbers $J^{PC}$ and the decay modes.
In contrast to $J/\psi\rightarrow\gamma K^{+}K^{-}\eta^{\prime}$, there is no background contamination for $J/\psi\rightarrow\gamma K^{0}_{S}K^{0}_{S}\eta^{\prime}$ from $J/\psi\rightarrow \pi^{0} K^{0}_{S}K^{0}_{S}\eta^{\prime}$ and $J/\psi\rightarrow K^{0}_{S}K^{0}_{S}\eta^{\prime}$, which are forbidden by exchange symmetry and $CP$ conservation.
Therefore, the $J/\psi\rightarrow\gamma K^{0}_{S}K^{0}_{S}\eta^{\prime}$ decay provides a clean environment for its $J^{PC}$  measurement with minimal background modeling uncertainties.
In this Letter, we report the first spin-parity determination of the $X(2370)$ in the decay $J/\psi\rightarrow\gamma K^{0}_{S}K^{0}_{S}\eta^{\prime}$, 
where the $K^{0}_{S}$ decays to $\pi^{+}\pi^{-}$
and the $\eta^{\prime}$ decays to the two most dominant channels $\eta^{\prime}\rightarrow\gamma\pi^{+}\pi^{-}$ and $\eta^{\prime}\rightarrow\eta\pi^{+}\pi^{-}(\eta\rightarrow\gamma\gamma)$.
The analysis is based on $(10087\pm44)\times 10^{6}$ $J/\psi$ events \cite{number} collected in the BESIII detector \cite{detector}.

A detailed description of the design and performance of the BESIII detector can be found in Ref.~\cite{detector}.
Simulated samples produced with a {\footnotesize GEANT4}-based \cite{geant4} Monte Carlo (MC) package, which includes the geometric description of the BESIII detector~\cite{detector2022} and the detector response, are used 
for the optimization of event selection criteria and detection efficiency determination.
Signal MC samples for the process $J/\psi \rightarrow \gamma K^{0}_{S}K^{0}_{S}\eta^{\prime}$ with the subsequent decays $K^{0}_{S} \rightarrow \pi^{+}\pi^{-}$, $\eta^{\prime} \rightarrow \pi^{+}\pi^{-}\eta$, and $\eta \to \gamma\gamma$ are generated uniformly in phase space.
A special generator takes $\rho - \omega$ interference and box anomaly into account \cite{etap_generator_2018} in the process of $\eta^{\prime}\rightarrow\gamma\pi^{+}\pi^{-}$.

Charged tracks reconstructed from the multilayer drift chamber (MDC) are required to be within the polar angle range $\vert\!\cos\theta\vert < 0.93$, 
where $\theta$ is defined with respect to the $z$ axis, which is the symmetry axis of the MDC.
The distance of closest approach to the interaction point for charged tracks (excluding those from $K_{S}^{0}$ decays) must be
less than 10 \textrm{cm} along the $z$ axis and less than 1 \textrm{cm} in the transverse plane. 
All charged tracks are assumed to be pions.
To reconstruct $K_{S}^{0}$ candidates, the tracks of each $\pi^{+}\pi^{-}$ pair are fitted to a secondary vertex.
To suppress background events, all $K_{S}^{0}$ candidates are required to satisfy $|M_{\pi^{+}\pi^{-}}-m_{K_{S}^{0}}|<9~\mathrm{MeV}/c^{2}$, where $m_{K_{S}^{0}}$ is the known mass of $K_{S}^{0}$~\cite{pdg}. 
To further suppress background, the decay length of $K_{S}^{0}$ candidate, i.e. the distance between the average position of the $e^{+}e^{-}$ collisions and the decay vertex of $K_{S}^{0}$, is required to be greater than twice the vertex resolution. 
With these selections, the miscombination of $K_{S}^{0}$ reconstruction is significantly suppressed to be less than 0.1\%.
The reconstructed $K_{S}^{0}$ candidates are used as an input for the subsequent kinematic fit.

Photon candidates are identified using showers in the electromagnetic calorimeter (EMC). 
The deposited energy of each shower are required to have at least 100 MeV in the barrel region ($\vert\!\cos\theta\vert<0.80$) and the end cap region ($0.86<\vert\!\cos\theta\vert<0.92$). 
To exclude showers from charged tracks, the angle between the shower position and the charged tracks extrapolated to the EMC must be greater than $10^{\circ}$. 
The difference between the EMC time and the event start time is required to be within [0, 700] ns in order to suppress electronic noise and energy deposits unrelated to the event.

For the $J/\psi\rightarrow\gamma K_{S}^{0}K_{S}^{0}\eta^{\prime}, \eta^{\prime}\rightarrow\gamma\pi^{+}\pi^{-}$ channel, 
each candidate event is required to have at least three positively charged tracks, at least three negatively charged tracks and two photons.
A four-constraint (4C) kinematic fit under the $J/\psi\rightarrow\gamma\gamma K_{S}^{0}K_{S}^{0}\pi^{+}\pi^{-}$ hypothesis is performed by enforcing energy-momentum conservation. 
If there is more than one $\gamma\gamma K_{S}^{0}K_{S}^{0}\pi^{+}\pi^{-}$ combination, the one with the smallest $\chi^{2}_{\rm 4C}$ is chosen. The resulting $\chi^{2}_{\rm 4C}$ is required to be less than 40. 
The $\eta^{\prime}$ candidates are required to have the invariant mass satisfying $|M_{\gamma\pi^{+}\pi^{-}} - m_{\eta^{\prime}}| < 15~\mathrm{MeV}/c^{2}$, 
where $m_{\eta^{\prime}}$ is the known mass of $\eta^{\prime}$ \cite{pdg}.
If there is more than one $\gamma\pi^{+}\pi^{-}$ combination, the one with the minimum $|M_{\gamma\pi^{+}\pi^{-}} - m_{\eta^{\prime}}|$ is selected. The $\pi^{+}\pi^{-}$ (from $\eta^{\prime}$) invariant mass is required to be in the $\rho$ mass region, $0.55 < M_{\pi^{+}\pi^{-}} <  0.90~\mathrm{GeV}/c^{2}$.
To suppress background events containing a $\pi^{0}$ or $\eta$, events with $|M_{\gamma\gamma} - m_{\pi^{0}}| < 20~\mathrm{MeV}/c^{2}$ or $|M_{\gamma\gamma} - m_{\eta}| < 30~\mathrm{MeV}/c^{2}$ are rejected, where $m_{\pi^{0}}$ and $m_{\eta}$ are the known masses of $\pi^{0}$ and $\eta$, respectively~\cite{pdg}.

For the $J/\psi\rightarrow\gamma K_{S}^{0}K_{S}^{0}\eta^{\prime}, \eta^{\prime}\rightarrow\pi^{+}\pi^{-}\eta,\eta\rightarrow\gamma\gamma$ channel, 
each candidate event is required to have at least three positively charged tracks, at least three negatively charged tracks and three photons.
A 4C kinematic fit is performed under the $J/\psi\rightarrow\gamma\gamma\gamma K_{S}^{0}K_{S}^{0}\pi^{+}\pi^{-}$ hypothesis and the combination with the smallest $\chi^{2}_{\rm 4C}$ is chosen if more than one combination is found.
In order to reduce background and to improve the mass resolution,  
a five-constraint (5C) kinematic fit is performed to further constrain 
the invariant mass of the two photons to $m_\eta$.
Among three $\gamma\gamma$ combinations, the one with the smallest $\chi^{2}_{\rm 5C}$ is chosen, and $\chi^{2}_{\rm 5C}<50$ is required. 
The $\eta^{\prime}$ candidates must satisfy $|M_{\pi^{+}\pi^{-}\eta} - m_{\eta^{\prime}}| < 10~\mathrm{MeV}/c^{2}$.
To suppress background events containing a $\pi^{0}$, events with $|M_{\gamma\gamma} - m_{\pi^{0}}| < 20~\mathrm{MeV}/c^{2}$ are rejected, where the photon pairs are all possible combinations of the radiative photon and photons from $\eta$.

All the above selection criteria aim to improve the signal extraction efficiency and signal-to-noise ratio. 
The mass windows for peaking signals of $K_{S}^{0}$ and $\eta^{\prime}$ correspond to approximately 3 standard deviations to their respective known masses~\cite{pdg}. 
Others are determined by optimizing the figure of merit (FOM) $\epsilon_\text{S}/\sqrt{\text{N}_{\text{data}}}$, where $\epsilon_\text{S}$ is signal efficiency with simulation MC sample, and $\text{N}_{\text{data}}$ is the final selected event number in data.
With above criteria, the event numbers of final selected candidates are 4046 and 1395 for the $\eta^{\prime}\rightarrow\gamma\pi^+\pi^-$ channel and the $\eta^{\prime}\rightarrow\pi^+\pi^-\eta$ channel, respectively.

No significant peaking background contribution has been found in the measured invariant mass spectra.
The remaining background component is from non-$\eta^{\prime}$ processes, 
which are estimated from the $\eta^{\prime}$ mass sideband regions of
$20 < |M_{\gamma\pi^{+}\pi^{-}} - m_{\eta^{\prime}}| < 30~\mathrm{MeV}/c^{2}$  
and $30 < |M_{\pi^{+}\pi^{-}\eta} - m_{\eta^{\prime}}| < 40~\mathrm{MeV}/c^{2}$.
The corresponding background fractions are $6.8\%$ and $1.8\%$ 
for the two channels, respectively. 

\begin{figure}[htbp]
\centering
\vspace{-2.0mm}

\hspace{-3.2mm}
\subfloat{
	\includegraphics[width=0.5\textwidth]{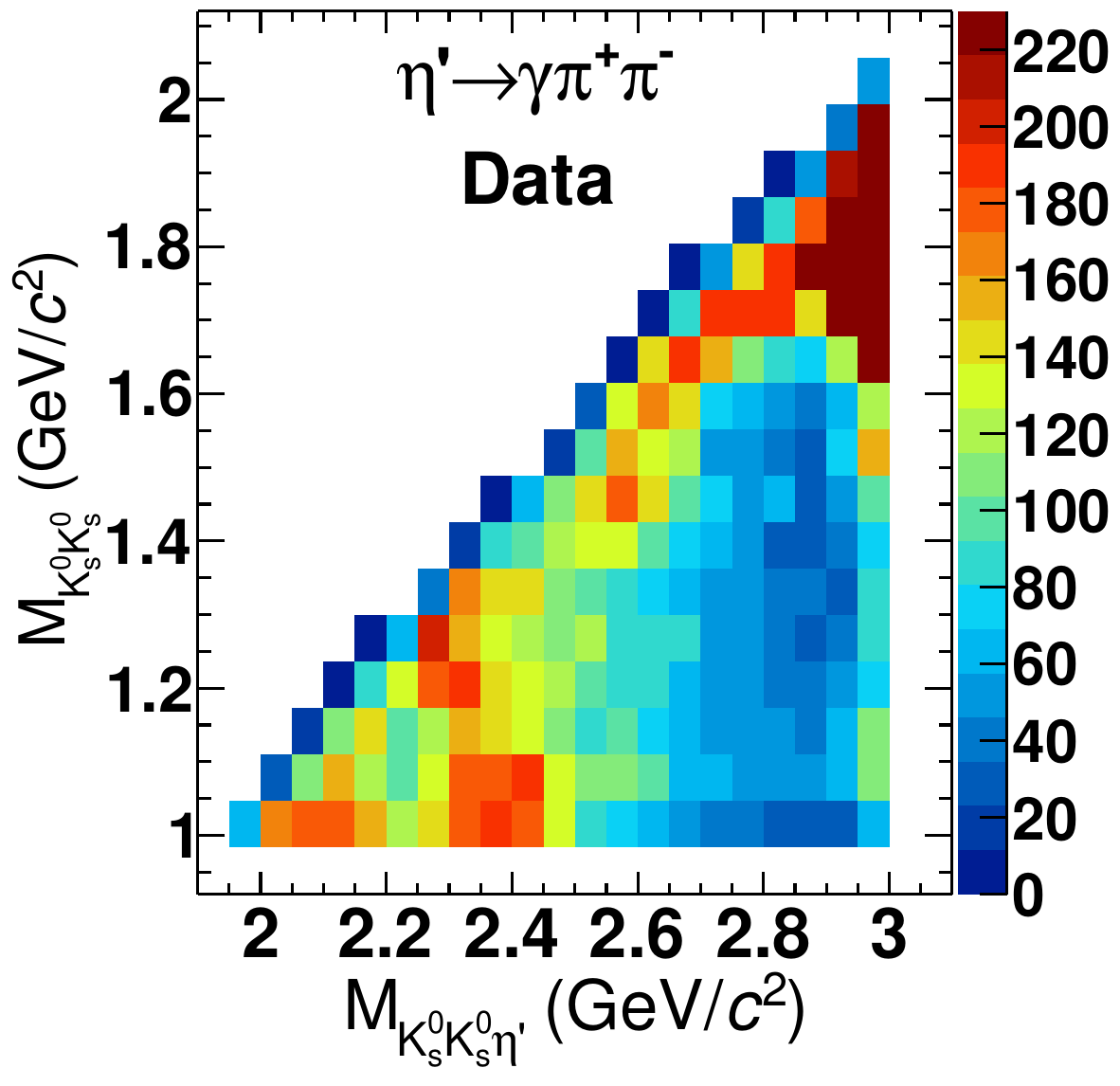}
	\label{fig:Subfigure1}
\put(-102,105){(a)}
}
\hspace{-3.2mm}
\subfloat{
	\includegraphics[width=0.5\textwidth]{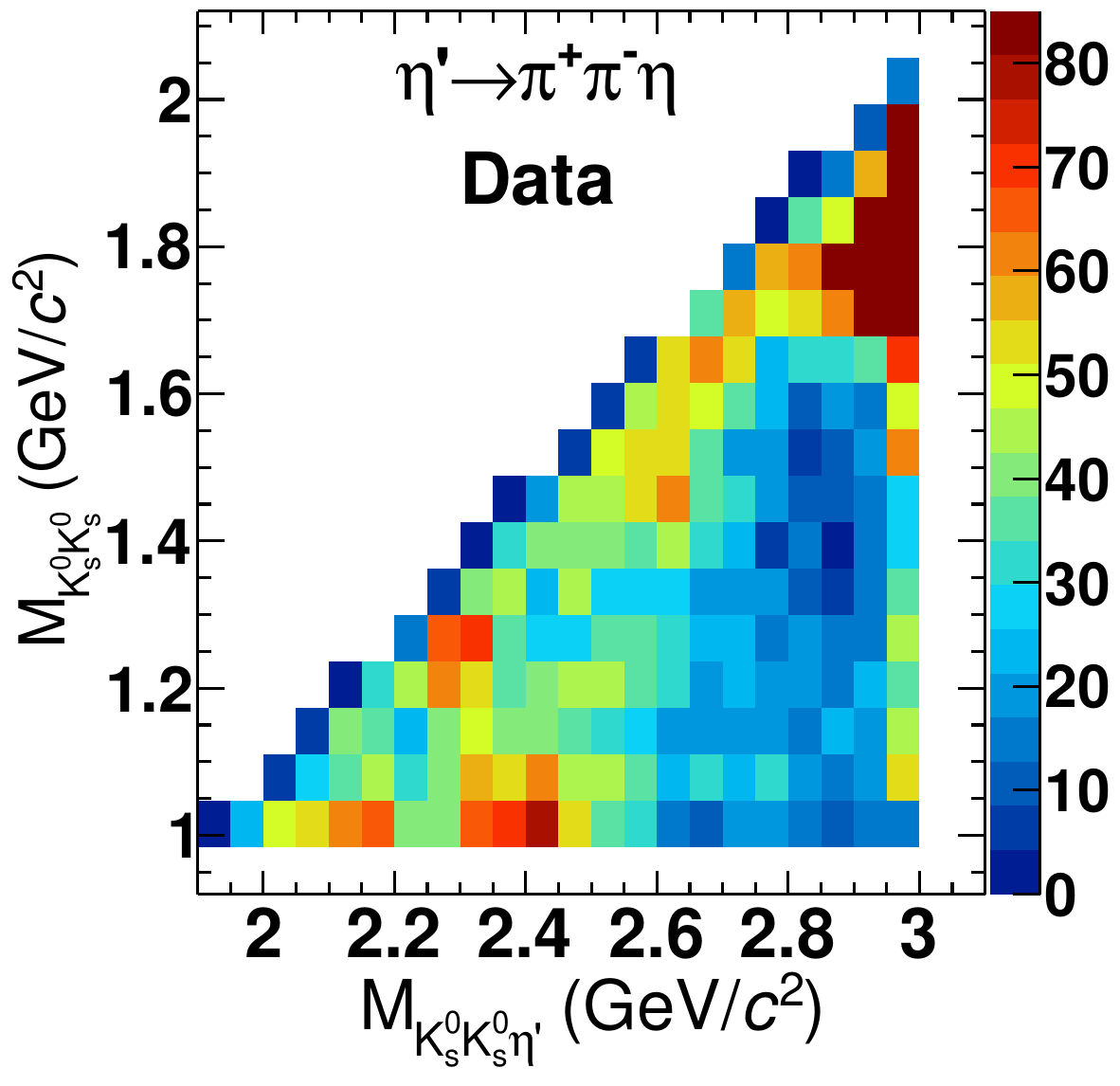}
	\label{fig:Subfigure2}
\put(-102,105){(b)}
}
\vspace{-3.0mm}

\hspace{-3.2mm}
\subfloat{
	\includegraphics[width=0.5\textwidth]{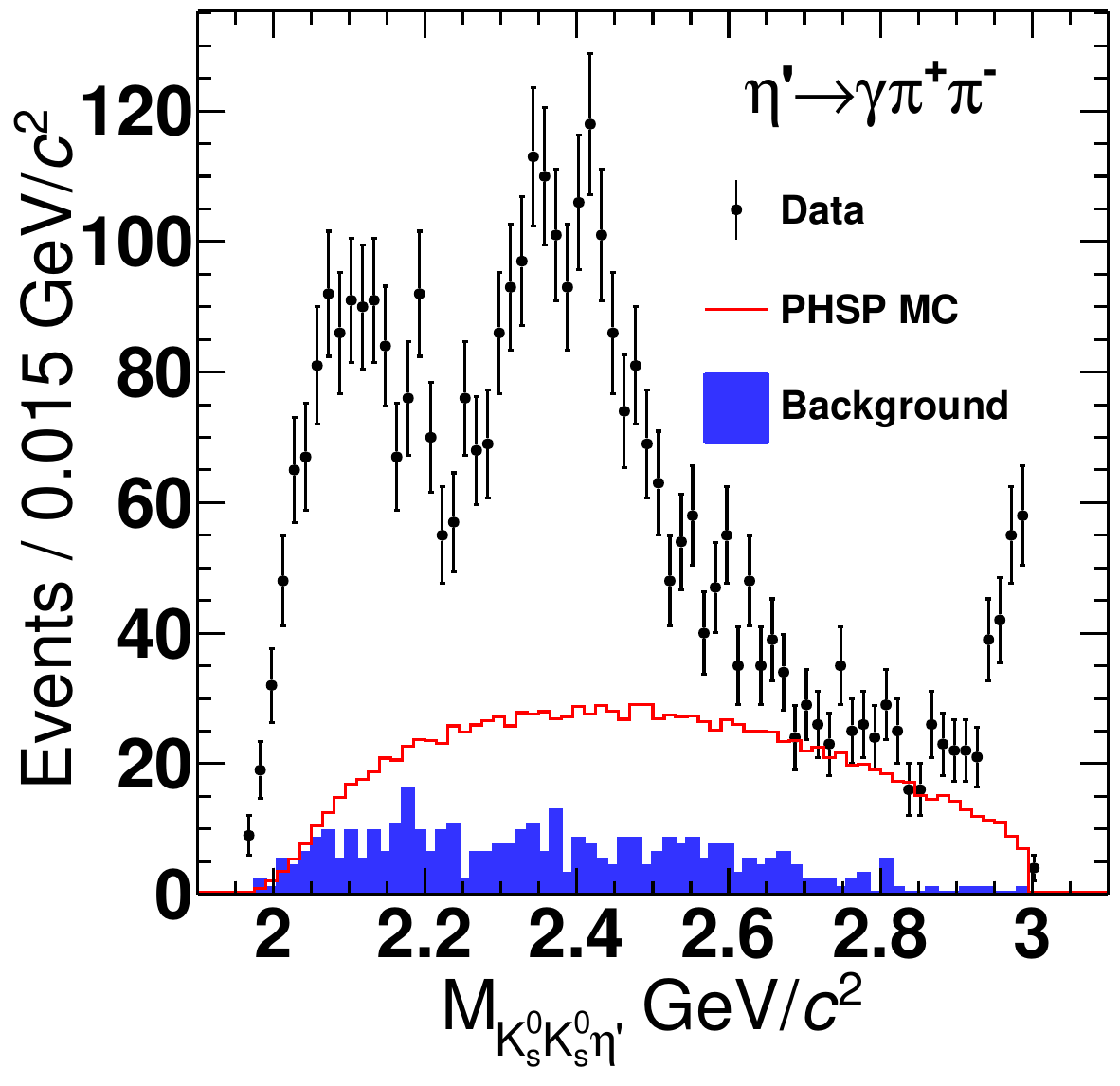}
	\label{fig:Subfigure3}	
\put(-102,105){(c)}
}
\hspace{-3.2mm}
\subfloat{
	\includegraphics[width=0.5\textwidth]{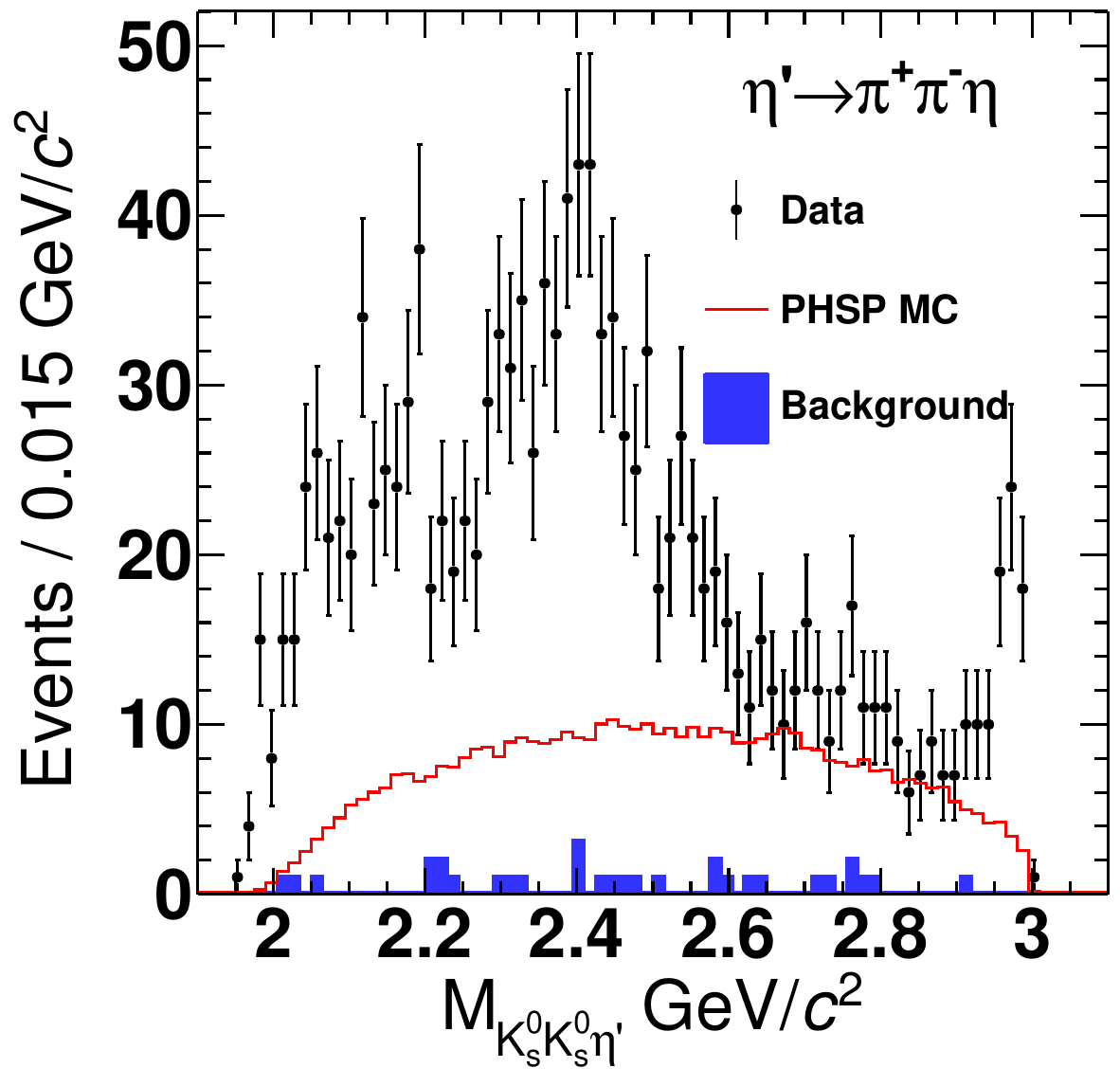}
	\label{fig:Subfigure4}
\put(-102,105){(d)}
}

\caption{
Invariant mass distributions of the selected events: (a) and (b) are the two-dimensional distributions of $M_{K_{S}^{0}K_{S}^{0}}$ versus $M_{K_{S}^{0}K_{S}^{0}\eta^{\prime}}$ for the $\eta^{\prime}\rightarrow\gamma\pi^{+}\pi^{-}$ and $\eta^{\prime}\rightarrow\pi^{+}\pi^{-}\eta$ channels, respectively.
(c) and (d) are the $K_{S}^{0}K_{S}^{0}\eta^{\prime}$ invariant mass distributions with the requirement $M_{K_{S}^{0}K_{S}^{0}}<1.1~\mathrm{GeV}/c^{2}$ for $\eta^{\prime}\rightarrow\gamma\pi^{+}\pi^{-}$ and $\eta^{\prime}\rightarrow\pi^{+}\pi^{-}\eta$ channels, respectively.
The dots with error bars are data. The shaded histograms are the non-$\eta^{\prime}$ backgrounds estimated by the $\eta^{\prime}$ sideband. The solid lines are phase space (PHSP) MC events with arbitrary normalization.
}

\label{fig:fig1}
\end{figure}

Figure~\ref{fig:fig1} shows the mass distributions with the above selection criteria 
for the $\eta^{\prime}\rightarrow\gamma\pi^+\pi^-$ and $\eta^{\prime}\rightarrow\pi^+\pi^-\eta$ channels.
Similar structures are observed in the two channels.
The two-dimensional distributions of $M_{K^{0}_{S}K^{0}_{S}}$ versus $M_{K^{0}_{S}K^{0}_{S}\eta^{\prime}}$ indicate a strong enhancement near the $K^{0}_{S}K^{0}_{S}$ mass threshold from the $f_{0}(980)$
and a clear connection between the $f_{0}(980)$ and the structure around $2.4\ \mathrm{GeV}/c^{2}$, $X(2370)$, in the invariant mass spectra of $K_{S}^{0}K_{S}^{0}\eta^{\prime}$.
By requiring $M_{K^{0}_{S}K^{0}_{S}} < 1.1\ \mathrm{GeV}/c^{2}$, the structure around $2.4\ \mathrm{GeV}/c^{2}$ becomes much more prominent 
in the $K_{S}^{0}K_{S}^{0}\eta^{\prime}$ mass spectrum. 
In addition, there is a clear signature from the $\eta_{c}$.

A partial wave analysis (PWA) is performed to investigate the properties of the $X(2370)$. To reduce complexities from additional intermediate processes, events satisfying $M_{K_{S}^{0}K_{S}^{0}}<1.1~\mathrm{GeV}/c^{2}$ are used.
The $K_{S}^{0}$ and $\eta^{\prime}$ momenta are constrained to their known masses, respectively. 
The signal amplitudes are constructed with the covariant tensor formalism \cite{zoubs2003} and parameterized as quasi-sequential two-body decays: $J/\psi\rightarrow \gamma X$, $X\rightarrow Y\eta^{\prime}$ or $X\rightarrow Z K^{0}_{S}$, where $Y$ and $Z$ represent $K^{0}_{S}K^{0}_{S}$ and $K^{0}_{S}\eta^{\prime}$ isobars, respectively. 
Due to the parity conservation, the possible $J^{PC}$ of $K^{0}_{S}K^{0}_{S}\eta^{\prime}$ system ($X$) are $0^{-+}$,$1^{++}$,$2^{++}$,$2^{-+}$, etc. 
In this Letter, given the suppression of phase space factor, only spin $J<3$ states of the $X$ and possible S-wave or P-wave and D-wave decays of intermediate states are considered.
An unbinned maximum likelihood fit is performed on the combined data of the two $\eta^{\prime}$ decay modes.
The non-$\eta^{\prime}$ background contribution is taken into account in the fit via the subtraction of the negative log-likelihood values 
with the events estimated from the $\eta^{\prime}$ mass sideband region.

\begin{figure}[htbp]
\centering
\vspace{-3.0mm}

\hspace{-4.0mm}
\subfloat{
	\includegraphics[width=0.5\textwidth]{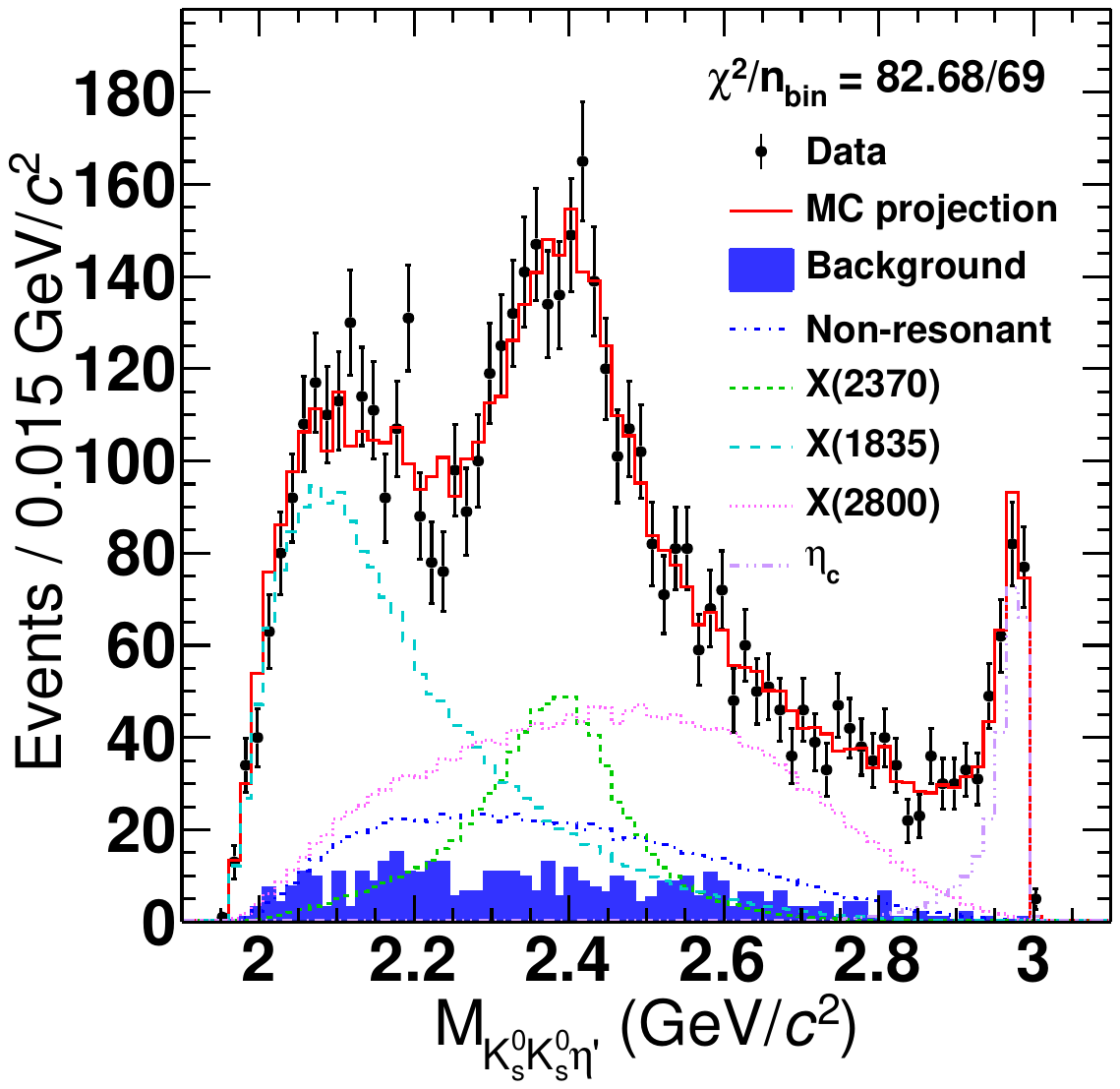}
	\label{fig:Subfigure11}
\put(-102,105){(a)}
}
\hspace{-4.0mm}
\subfloat{
	\includegraphics[width=0.5\textwidth]{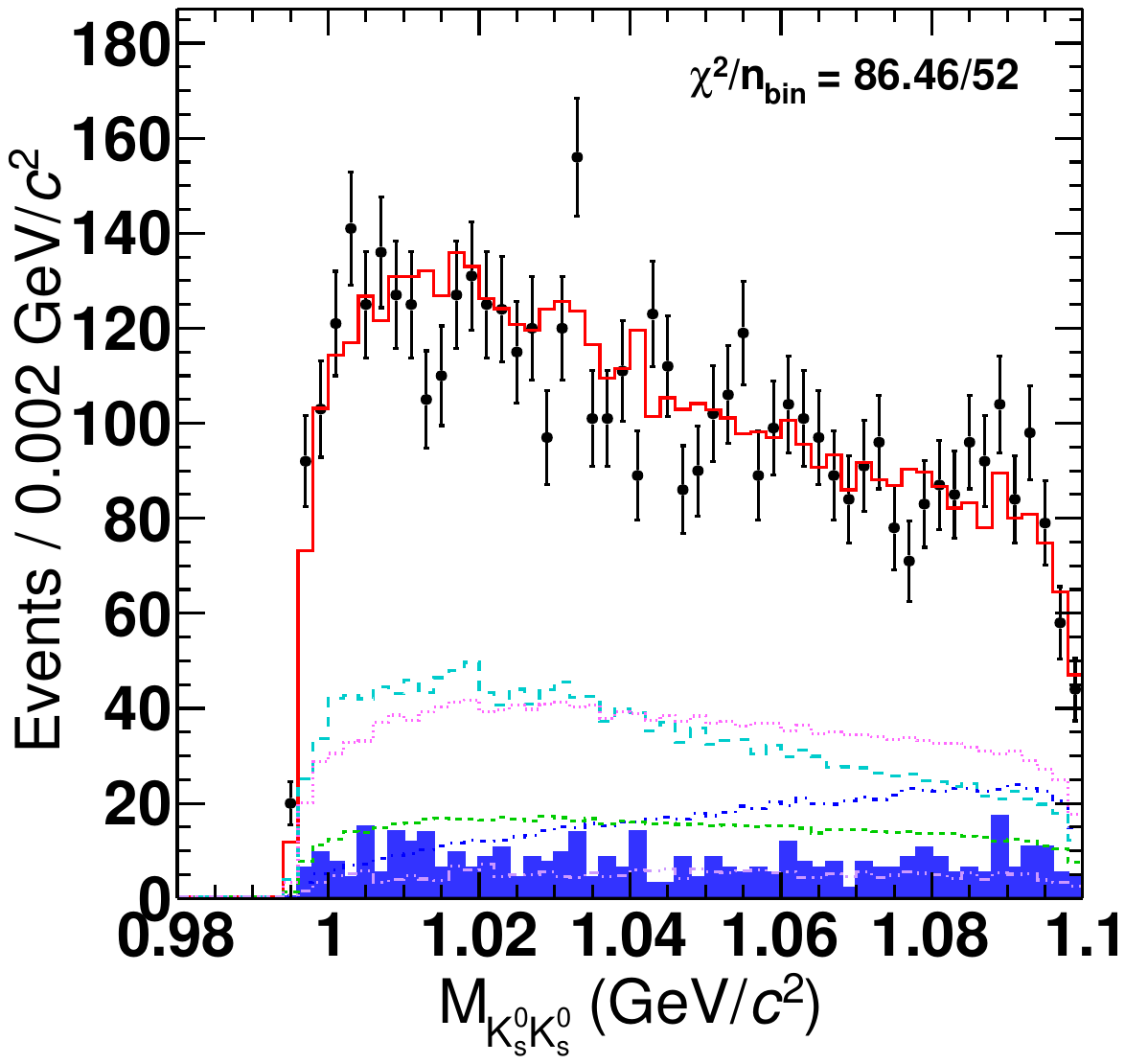}
	\label{fig:Subfigure12}
\put(-102,105){(b)}
}
\vspace{-3.0mm}

\hspace{-4.0mm}
\subfloat{
	\includegraphics[width=0.5\textwidth]{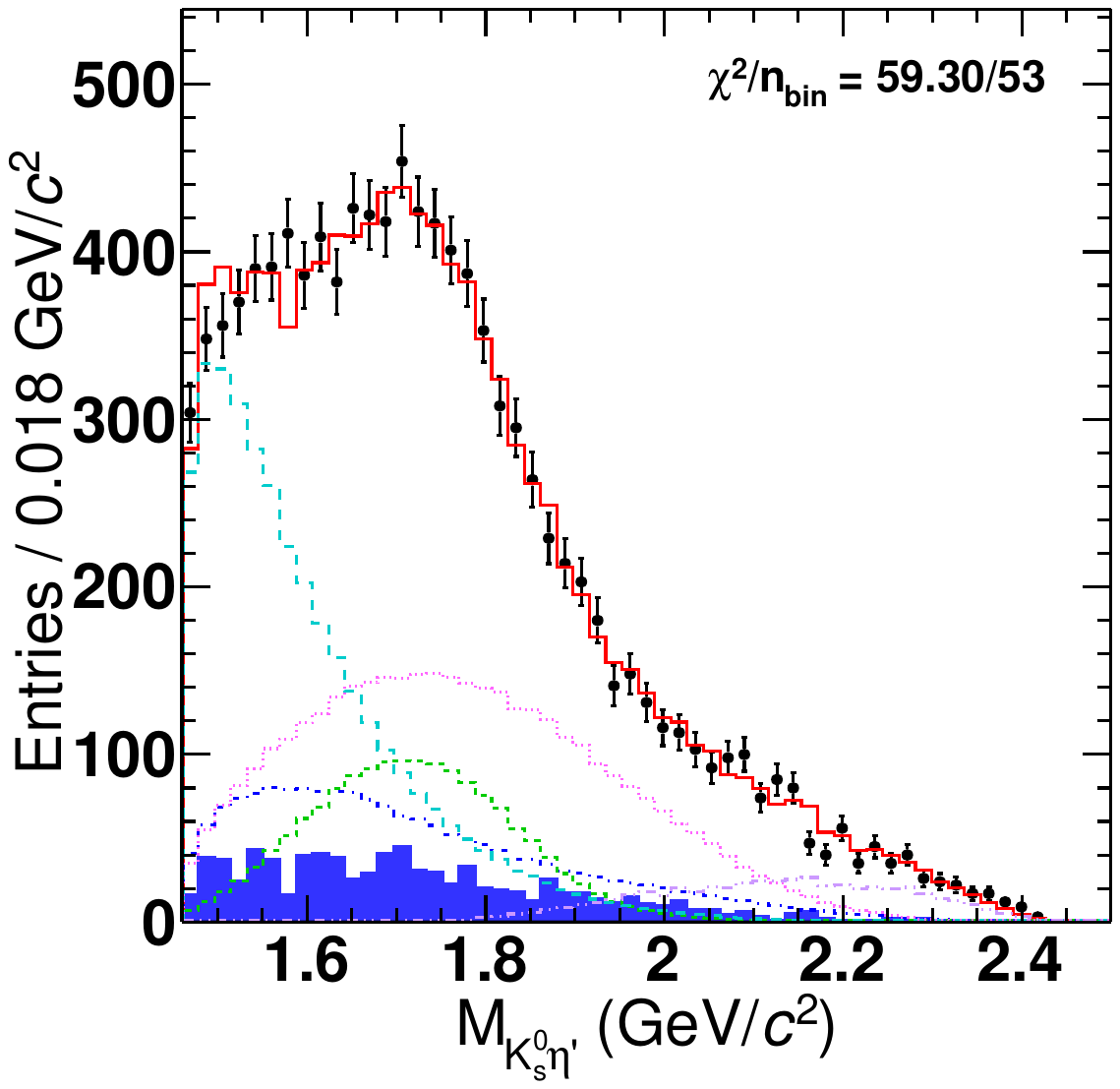}
	\label{fig:Subfigure13}	
\put(-102,105){(c)}
}
\hspace{-4.0mm}
\subfloat{
	\includegraphics[width=0.5\textwidth]{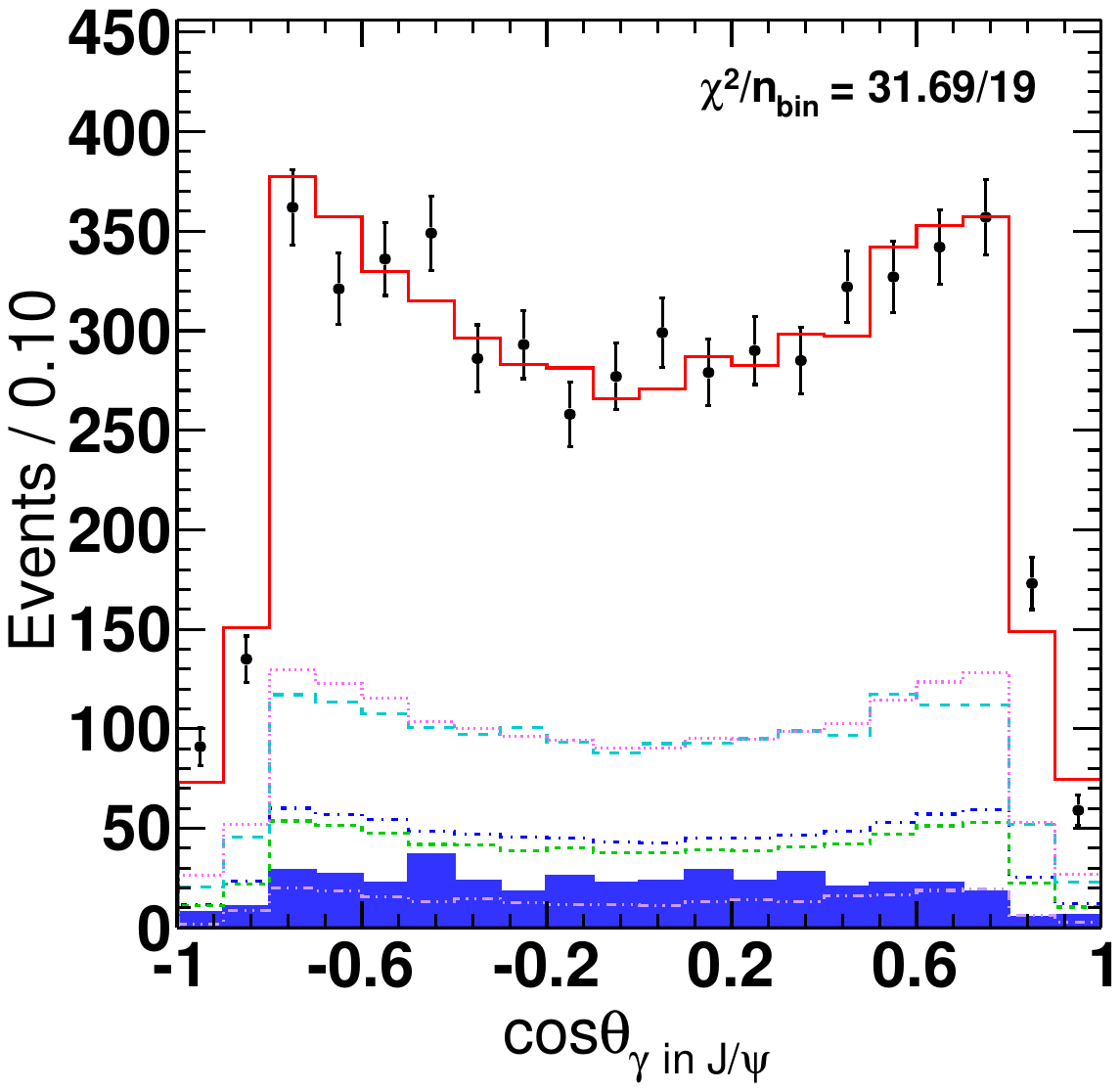}
	\label{fig:Subfigure14}
\put(-102,105){(d)}
}
\vspace{-3.0mm}

\hspace{-4.0mm}
\subfloat{
	\includegraphics[width=0.5\textwidth]{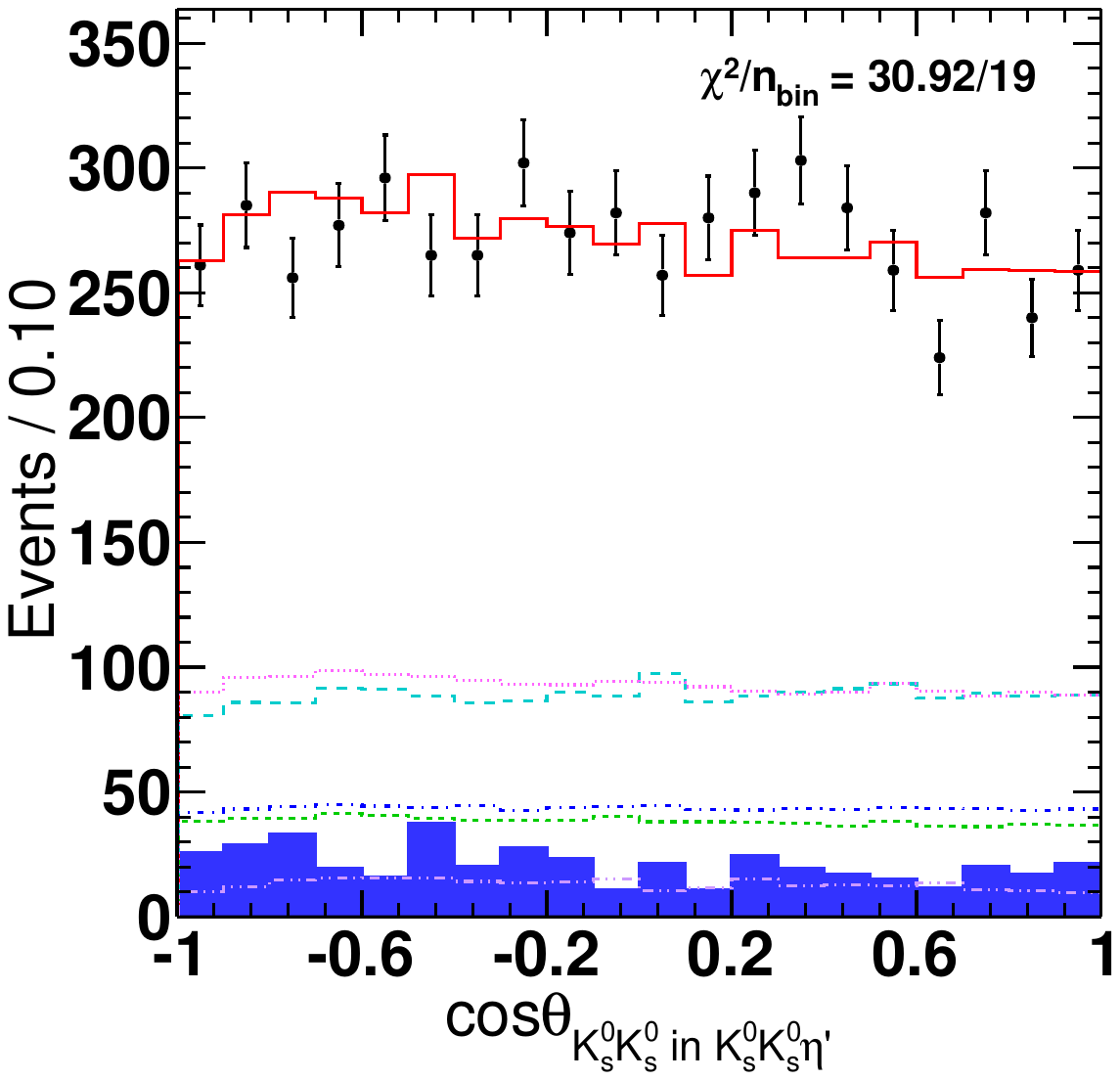}
	\label{fig:Subfigure11}
\put(-102,105){(e)}
}
\hspace{-4.0mm}
\subfloat{
	\includegraphics[width=0.5\textwidth]{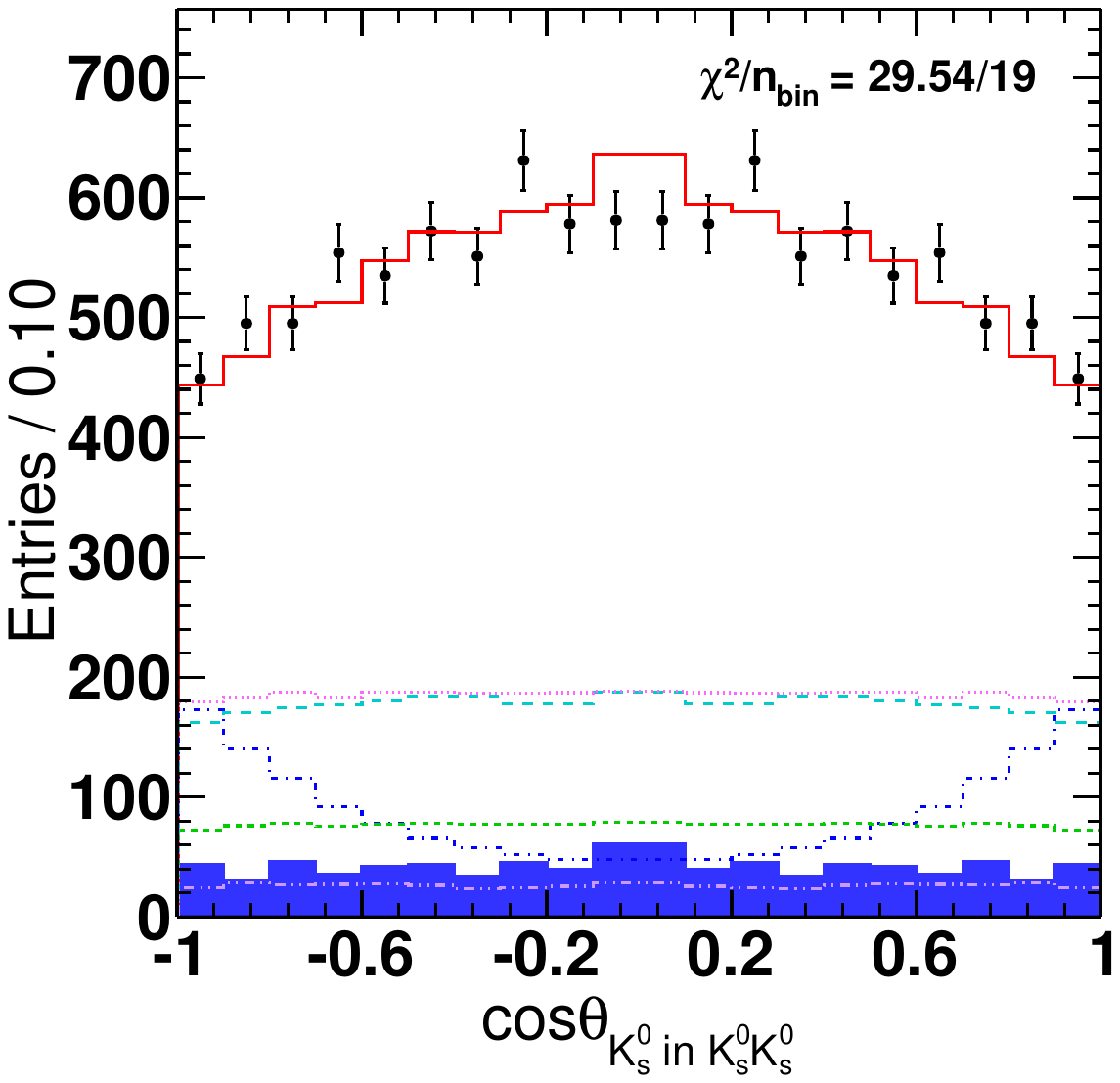}
	\label{fig:Subfigure12}
\put(-102,105){(f)}
}

\vspace{-1.5mm}
 \caption{
  Comparisons between data (with two $\eta^{\prime}$ decay modes combined) and PWA fit projections: \textcolor{blue}{(a)}, \textcolor{blue}{(b)}, and \textcolor{blue}{(c)} are the invariant mass distributions of $K^{0}_{S}K^{0}_{S}\eta^{\prime}$, $K^{0}_{S}K^{0}_{S}$, and $K^{0}_{S}\eta$ (two entries for one event), respectively. 
  \textcolor{blue}{(d)}, \textcolor{blue}{(e)} and \textcolor{blue}{(f)} are the angular distributions of $\cos\theta$, where $\theta$ is the polar angle of (d) $\gamma$ in the $J/\psi$ rest system;
  (e) $K^{0}_{S}K^{0}_{S}$ in the $K^{0}_{S}K^{0}_{S}\eta^{\prime}$ rest system; and (f) $K^{0}_{S}$ in the $K^{0}_{S}K^{0}_{S}$ rest system (two entries for one event). The dots with error bars are data. The solid red histograms are the PWA total projections. The shaded histograms are the non-$\eta^{\prime}$ backgrounds described by the $\eta^{\prime}$ sideband. The dash-dotted blue, short dashed green, long dashed cyan, dotted magenta and dash-dot-dotted violet show the contributions of the non-resonant contribution, $X(2370)$, $X(1835)$, $X(2800)$ and $\eta_{c}$, respectively. 
  }

  \label{fig:pwa}
\end{figure}

The optimal PWA fit shows that data can be well described with a process combination of the decay of $f_{0}(980)\eta^{\prime}$ from the resonances of the $X(1835)$, $X(2370)$, $\eta_{c}$ and a broad $0^{-+}$ structure denoted as $X(2800)$,
and the non-resonance components of $(K^{0}_{S}K^{0}_{S})_{S}\eta^{\prime}$ and $(K^{0}_{S}K^{0}_{S})_{D}\eta^{\prime}$ 
for the S-wave and D-wave in the $K^{0}_{S}K^{0}_{S}$ system, respectively.
The $X(1835)$, $X(2370)$ and $X(2800)$ are described 
by nonrelativistic Breit-Wigner (BW) functions, where the intrinsic widths are not energy dependent.
The masses and widths of the $X(1835)$ and $\eta_{c}$ are fixed to previous measurements \cite{jpc1835,pdg}.
The masses and widths of the $X(2370)$ and $X(2800)$ are floated in the PWA fit.
The mass line shape of $f_{0}(980)$ is parameterized by the Flatt\'e formula \cite{flatte1976} with the BESII measurement \cite{980resonances2005}. The $J^{PC}$ of the $X(2370)$ and $X(2800)$ are assigned to be $0^{-+}$.
The statistical significance of the $X(2370)$ is greater than $11.7 \sigma$, 
which is determined from the changes of log-likelihood value and degrees of freedom in the PWA fits with and without the signal hypotheses for every systematic variation.
The mass, width and product branching fraction of $X(2370)$ are measured to be
$2395 \pm 11 ({\rm stat})~\mathrm{MeV}/c^{2}$, $188^{+18}_{-17}({\rm stat})~\mathrm{MeV}/c^{2}$ 
and $\mathcal{B}[J/\psi\rightarrow\gamma X(2370)] \times \mathcal{B}[X(2370) \rightarrow f_{0}(980)\eta^{\prime}] \times \mathcal{B}[f_{0}(980) \rightarrow K^{0}_{S}K^{0}_{S}] = \left( 1.31 \pm 0.22 ({\rm stat})\right) \times 10^{-5}$, respectively.
Figure~\ref{fig:pwa} provides the comparisons of the mass and angular distributions between data and PWA fit projections, as well as the individual contributions from each component.
The $\chi^2\mathrm{/n_{bin}}$ value is displayed on each figure to demonstrate the goodness of fit.
A broad $0^{-+}$ structure is needed in the optimal PWA fit to describe the effective contributions from possible high-mass resonances such as $X(2600)$ \cite{2600para} 
and the tail of $\eta_{c}$ line shape, which is denoted as $X(2800)$ (with a mass of $2799~\mathrm{MeV}/c^{2}$ and a width of $660~\mathrm{MeV}/c^{2}$).
The $X(2800)$ have been checked with various alternative PWA fits. For example, if the $\eta_{c}$ line shape is parameterized without a damping factor \cite{etac_measurement}, the significance of $X(2800)$ is reduced to $3.1 \sigma$. If the $X(2800)$ is not included in the PWA, the spin-parity of $X(2370)$ remains to be $0^{-+}$ with a significance greater than $10.1 \sigma$. 
The significance of $0^{-+}$ over other alternative $J^{PC}$ is determined from the changes of log-likelihood value and degrees of freedom in PWA fits.
The impacts of the $X(2800)$ on the mass, width and product branching fraction of the $X(2370)$ are included in the systematic uncertainties.


Variations of the PWA fit including the $J^{PC}$ and decay mode for each component are tested.
Possible decay modes 
[$f_{0}(1500)\eta^{\prime}$, $f_{2}(1270)\eta^{\prime}$, $K^{*}(1410)K^{0}_{S}$, $K^{*}_{0}(1430) K^{0}_{S}$, $K^{*}_{2}(1430) K^{0}_{S}$, $K^{*}(1680)K^{0}_{S}$, $(K^{0}_{S}K^{0}_{S})_{S}\eta^{\prime}$, $(K^{0}_{S}K^{0}_{S})_{D}\eta^{\prime}$, $(K^{0}_{S}\eta^{\prime})_{P}K^{0}_{S}$, $(K^{0}_{S}\eta^{\prime})_{D}K^{0}_{S}$] 
are evaluated via different process combinations.
All additional decay modes have significances lower than $3\sigma$.
The contributions from additional resonances are also evaluated, including the $\eta(1760)$, $\eta(2225)$, $\eta_{2}(1870)$, $X(2120)$ \cite{1835_confirmed} and $X(2600)$ \cite{2600para}. 
All the significances of each contribution are measured to be less than $3\sigma$, except the $X(2600)$. The significance of the process of  $X(2600)\rightarrow f_{0}(980)\eta^{\prime}$ is $4.2\sigma$. This process is not included in the optimal solution, but the possible contribution of this process is taken into account as a source of systematic uncertainties.
The scan results yield no evidence for extra intermediate states.
For the spin-parity determination of the $X(2370)$, the $0^{-+}$ assignment fit is better than that for $1^{++}$ or $2^{-+}$ assignments with significances that are greater than $10.8\sigma$ or $9.8\sigma$, respectively.
The significances are evaluated with the consideration of all systematic uncertainty variations as described below.

Systematic uncertainty associated with the PWA affects both the branching fraction measurement and the resonance parameters,
including the background contribution, $f_{0}(980)$ mass line shape, the $X(1835)$ mass line shape, $\eta_c$ mass line shape, BW formula, additional resonances and description of the broad $0^{-+}$ structure. 
The uncertainty due to the background contribution is estimated using different background normalization factors and different $\eta^{\prime}$ sideband regions.
The $f_{0}(980)$ mass line shape is varied by changing the mass and coupling constants in the $Flatt\acute{e}$ formula to other experimental measurements \cite{980sys_LHCb2012}. 
Uncertainty from the $X(1835)$ mass line shape includes the variation with 1 standard deviation of the mass and width measurement \cite{jpc1835} and the alternative parametrization of the anomalous line shape  near the $p\bar{p}$ mass threshold \cite{1835parameterization}.  
Uncertainty from the $\eta_c$ mass line shape is estimated
by turning off the damping factor \cite{etac_measurement}.
Uncertainty arising from the BW parametrization is estimated by replacing the constant width with a mass-dependence width \cite{mass_dependent_width}. 
The impact from possible additional resonances is estimated by including the contributions of $X(2120)$ and $X(2600)$ to the PWA fit.
The broad $0^{-+}$ structure is described with the $X(2800)$ in the optimal PWA fit and has been checked with various PWA fits including replacing the $X(2800)$ with a non-resonance component of $f_{0}(980)\eta^\prime$, removing the $X(2800)$ and adding a non-resonance component of $f_{0}(980)\eta^\prime$ for the exclusion of the damping factor for the $\eta_c$.
The envelope of those variations is assigned as the final uncertainty from the description of the broad $0^{-+}$ structure. This is the dominant systematic uncertainty source for the measurements of mass, width and product branching fraction of the $X(2370)$.

Additional systematic uncertainty associated with the event selection, including tracking efficiency~\cite{tracking_uncertainty}, photon selection efficiency~\cite{PhysRevD.81.052005}, 
kinematic fit~\cite{helix_method}, 
$K^{0}_{S}$ reconstruction~\cite{ks_uncertainty},
the branching fractions of $K^{0}_{S}\rightarrow\pi^{+}\pi^{-}$, $\eta^{\prime}\rightarrow\pi^{+}\pi^{-}\eta$, $\eta^{\prime}\rightarrow\gamma\pi^{+}\pi^{-}$ and $\eta\rightarrow\gamma\gamma$~\cite{pdg},
and the total number of $J/\psi$ events \cite{number}, has been estimated to be $\pm4.8\%$ for the measurement of product branching fraction.
All studied systematic uncertainty sources and their contributions are summarized in Table~\ref{tab:sys} and are treated independently. 
Total systematic uncertainties on the mass and width of the $X(2370)$ are $^{+26}_{-94}~\mathrm{MeV}/c^2$ and $^{+124}_{-33}~\mathrm{MeV}$, respectively,
and total relative systematic uncertainty on the corresponding product branching fraction is $^{+217.0}_{-63.7}\%$.

In summary, a PWA of $J/\psi\rightarrow\gamma K^{0}_{S}K^{0}_{S}\eta^{\prime}$ has been performed in the full $K^{0}_{S}K^{0}_{S}\eta^{\prime}$ invariant mass range with the requirement of $M_{K^{0}_{S}K^{0}_{S}}<1.1~\mathrm{GeV}/c^{2}$. 
The PWA fit indicates a contribution from $X(2370)\rightarrow K^{0}_{S}K^{0}_{S}\eta^{\prime}$ with a statistical significance greater than 14$\sigma$.
The mass and width of the $X(2370)$ are measured to be $2395 \pm 11 ({\rm stat})^{+26}_{-94}({\rm syst})\ \mathrm{MeV}/c^{2}$ and $188^{+18}_{-17}({\rm stat})^{+124}_{-33}({\rm syst})~\mathrm{MeV}$, respectively. 
These results agree with the previous measurements from $J/\psi\rightarrow\gamma\pi^{+}\pi^{-}\eta^{\prime}$~\cite{1835_confirmed} and $J/\psi\rightarrow\gamma K\bar{K}\eta^{\prime}$~\cite{epjc.s10052_gongli}. 
The corresponding product branching fraction is $\mathcal{B}[J/\psi\rightarrow\gamma X(2370)] \times \mathcal{B}[X(2370) \rightarrow f_{0}(980)\eta^{\prime}] \times \mathcal{B}[f_{0}(980) \rightarrow K^{0}_{S}K^{0}_{S}] =  \left( 1.31 \pm 0.22 ({\rm stat})^{+2.85}_{-0.84}({\rm syst}) \right) \times 10^{-5}$.
The spin-parity of the $X(2370)$ is determined to be $0^{-+}$ for the first time. 
The measured mass of $X(2370)$ is in a good agreement with the mass prediction of the lightest pseudoscalar glueball, which is expected to be $(2.395\pm0.014)~\mathrm{GeV}/c^{2}$ from latest LQCD calculations \cite{LQCD5}.

\begin{table}[htp]
\centering
\renewcommand\arraystretch{1.2}
\caption{Systematic uncertainties on the measurements of mass, width and product branching fraction of the X(2370).
}

\resizebox{\columnwidth}{!}{

    \begin{tabular}{c|c|c|c}
    \hline\hline
    Sources & $\Delta M~(\mathrm{MeV}/c^{2})$ & $\Delta \Gamma~(\mathrm{MeV})$ & $\Delta \mathcal{B}/\mathcal{B}\ (\%)$ \\
    \hline

    Event selection                 & ---    & ---    & $\pm 4.8$ \\
    \hline
    Background estimation      		& $+2$ & $^{+4}_{-4}$ & $^{+3.7}_{-5.1}$ \\
    \hline
    $f_{0}(980)$ parameterization   & $-6$   & $+7$    & $\pm5.3$ \\
    \hline
    X(1835) parameterization 		& $^{+15}_{-12}$  & $^{+24}_{-11}$   & $^{+20.2}_{-8.3}$ \\
    \hline
    $\eta_{c}$ parameterization    	& $-13$   & $-8$    & $-14.5$ \\
    \hline
     Breit-Wigner formula     	& $-1$ & $+6$   & $-8.3$ \\
    \hline
    Broad $0^{-+}$ structure       		& $-88$   & $^{+111}_{-21}$   & $^{+211.8}_{-56.5}$ \\
    \hline    
    Additional resonances & $^{+22}_{-25}$   & $^{+48}_{-21}$   & $^{+41.9}_{-20.8}$ \\
    
    \hline
    Total                           & $^{+26}_{-94}$   & $^{+124}_{-33}$  & $^{+217.0}_{-63.7}$ \\ 

    \hline\hline
    \end{tabular}
    }

\label{tab:sys}
\end{table}

The BESIII Collaboration thanks the staff of BEPCII and the IHEP computing center for their strong support. This work is supported in part by National Key R\&D Program of China under Contracts Nos. 2020YFA0406300, 2020YFA0406400; National Natural Science Foundation of China (NSFC) under Contracts Nos. 11635010, 11735014, 11835012, 11935015, 11935016, 11935018, 11961141012, 12025502, 12035009, 12035013, 12061131003, 12192260, 12192261, 12192262, 12192263, 12192264, 12192265, 12221005, 12225509, 12235017; the Chinese Academy of Sciences (CAS) Large-Scale Scientific Facility Program; the CAS Center for Excellence in Particle Physics (CCEPP); Joint Large-Scale Scientific Facility Funds of the NSFC and CAS under Contract No. U1832207; CAS Key Research Program of Frontier Sciences under Contracts Nos. QYZDJ-SSW-SLH003, QYZDJ-SSW-SLH040; 100 Talents Program of CAS; The Institute of Nuclear and Particle Physics (INPAC) and Shanghai Key Laboratory for Particle Physics and Cosmology; European Union's Horizon 2020 research and innovation programme under Marie Sklodowska-Curie grant agreement under Contract No. 894790; German Research Foundation DFG under Contracts Nos. 455635585, Collaborative Research Center CRC 1044, FOR5327, GRK 2149; Istituto Nazionale di Fisica Nucleare, Italy; Ministry of Development of Turkey under Contract No. DPT2006K-120470; National Research Foundation of Korea under Contract No. NRF-2022R1A2C1092335; National Science and Technology fund of Mongolia; National Science Research and Innovation Fund (NSRF) via the Program Management Unit for Human Resources \& Institutional Development, Research and Innovation of Thailand under Contract No. B16F640076; Polish National Science Centre under Contract No. 2019/35/O/ST2/02907; The Swedish Research Council; U. S. Department of Energy under Contract No. DE-FG02-05ER41374.
\vspace{-0.2cm}

\bibliographystyle{apsrev4-1}
\bibliography{draft}

\end{document}